\documentclass[twocolumn,pra, twocolumn,showpacs,superscriptaddress, nofootinbib]{revtex4-1}

\usepackage[T1]{fontenc}
\usepackage[latin9]{inputenc}
\usepackage{color}
\usepackage{amsmath}
\usepackage{amssymb}
\usepackage{graphicx}
\usepackage[unicode=true,pdfusetitle,
 bookmarks=true,bookmarksnumbered=false,bookmarksopen=false,
 breaklinks=false,pdfborder={0 0 0},pdfborderstyle={},backref=false,colorlinks=true]
 {hyperref}
\hypersetup{
 citecolor=blue,urlcolor=blue}

\makeatletter
\usepackage{bbm}

\makeatother

\begin{document}
\title{The field-induced interaction between non-resonant magnetic dipoles}
\author{Fang-qi Hu}
\affiliation{Center for quantum technology research, School of Physics, Beijing
Institute of Technology, Beijing 100081, People's Republic of China}
\affiliation{China Academy of Engineering Physics, Beijing 100088, People's Republic
of China}
\author{Qing Zhao}
\email{qzhaoyuping@bit.edu.cn}

\affiliation{Center for quantum technology research, School of Physics, Beijing
Institute of Technology, Beijing 100081, People's Republic of China}
\author{Sheng-Wen Li}
\email{lishengwen@bit.edu.cn}

\affiliation{Center for quantum technology research, School of Physics, Beijing
Institute of Technology, Beijing 100081, People's Republic of China}
\begin{abstract}
We make a general derivation for the magnetic dipole-dipole interaction
based on the mediation of the quantized electro-magnetic field. Due
to the interaction with the dipoles, the dynamics of the field is
added by a dipole field, which finally gives rise to the dipole-dipole
interaction. Different from previous studies, the rotating-wave-approximation
is no longer needed throughout this derivation, and our result naturally
gives the interaction for non-resonant dipoles. Moreover, our derivation
also gives the counter-rotating interaction terms, and even the mixed
interaction terms between the permanent and transition dipoles. We
notice that this field-induced interaction is associated with the
interference of the virtual/real photons emitted from the two dipoles,
thus the interaction strength could be influenced by the frequency
difference of the two dipoles.
\end{abstract}
\maketitle

\section{Introduction}

The electric/magnetic dipole-dipole (DD) interaction exists in many
different physical systems \citep{oh2008entanglement,hu2018effect,balasubramanian2008nanoscale,pervushin1997attenuated,yang2010dimerization,el2013resonant},
such as the electric interaction between the dielectric defects in
solid-state systems \citep{martinis2005decoherence,paik2011observation,rigetti2012superconducting,lisenfeld_decoherence_2016,martinis_decoherence_2005},
and the magnetic interaction between the nitrogen-vacancy and the
nuclear spins around \citep{zhao2012decoherence,vznidarivc2008many,doherty_nitrogen-vacancy_2013,wrachtrup_processing_2006}.

In classical electrodynamics, it is well-known that the magnetic dipole-dipole
interaction is \citep{jackson_classical_1998}
\begin{equation}
V=\frac{\mu_{0}}{4\pi r^{3}}\big[\vec{m}_{1}\cdot\vec{m}_{2}-3(\vec{m}_{1}\cdot\hat{\mathrm{e}}_{\mathbf{r}})(\vec{m}_{2}\cdot\hat{\mathrm{e}}_{\mathbf{r}})\big].\label{eq:classical}
\end{equation}
For quantum systems, usually the DD interaction is obtained by simply
replacing the classical dipole moment $\vec{m}_{i}$ with the quantum
operator $\hat{\mathfrak{m}}_{i}$.

But this interaction also can be considered in another way: in principle,
the DD interaction should be induced by the mediation of the EM field.
To carry out this idea, we can derive a Markovian master equation
for two dipoles which are separated by a certain distance and locally
interacting with the EM field, then the field-induced DD interaction
appears as the off-diagonal Lamb shift in the unitary correction term
\citep{lehmberg1970radiation,agarwal_quantum_1974,ficek1987quantum,craig_molecular_1998,ficek2005quantum}. 

For two-level dipoles, this derivation gave a transition interaction
with the form $\hat{V}=g(\hat{\sigma}_{1}^{+}\hat{\sigma}_{2}^{-}+\hat{\sigma}_{1}^{-}\hat{\sigma}_{2}^{+})$
when their transition frequencies were exactly resonant. In this lengthy
derivation, the rotating-wave-approximation (RWA) should be applied
in a proper way, and the counter-rotating interaction terms cannot
be obtained, which is different from the above direct exchange method
from Eq.\,(\ref{eq:classical}).

Even for this resonant transition interaction, it turned out the interaction
strength $g$ does not exactly return to the above classical form
(\ref{eq:classical}), unless the distance between the two dipoles
is much smaller than the transition wavelength $r/\lambda\ll1$ (the
Dicke limit) \citep{lehmberg1970radiation,agarwal_quantum_1974,ficek1987quantum,craig_molecular_1998,ficek2005quantum,dicke1954coherence}. 

We noticed that this difference can be understood if the permanent
and transition dipoles are treated separately \citep{wang2018magnetic}.
By checking the dynamics of the EM field, we made another simplified
way to derive the field-induced magnetic DD interaction, which gives
both the permanent and transition DD interactions. The diagonal terms
of the dipole operator correspond to the permanent dipole moments,
and their interaction returns to the classical form directly. The
off-diagonal terms are the transition dipole moments, which indeed
do not have classical counterparts, thus their interaction intrinsically
contains quantum corrections and does not need to return to the classical
form exactly.

To make a more precise comparison with the interaction directly exchanged
from the classical one, we still need to study the DD interaction
between non-resonant dipoles, as well as the counter-rotating terms.
In this paper, we make a more general derivation for the field-induced
interaction between two magnetic dipoles with multiple energy levels.
We calculate the dynamics of the quantized EM field, and it becomes
the summation of the original vacuum field and the dipole field due
to the interaction with the dipoles. Then the DD interaction is derived
from the local interaction between the magnetic dipole and the incoming
dipole field. The RWA is no longer needed through out this derivation,
thus we can obtain the field-induced interaction for non-resonant
dipoles, and even the counter-rotating interaction terms.

Different from the electric interactions, both the static and radiative
magnetic interactions are related to the transverse modes of the EM
field, and do not involve the longitudinal modes \citep{agarwal_quantum_1974,cohen-tannoudji_photons_1989,craig_molecular_1998}.
Therefore, our derivations can be well done under the Coulomb gauge
and naturally gives both the permanent and transition DD interactions
between two non-resonant dipoles. Moreover, our result also contains
counter-rotating interaction terms and the mixed interaction terms
between the permanent and transition dipoles, which are usually negligible
under the RWA.

We notice that such field-induced DD interaction is associated with
the interference of the virtual/real photons emitted from the two
dipoles. As a result, the interaction strength could be influenced
by the the frequency difference of the two dipoles.

This paper is organized as follows. In Sec.\,\ref{sec2}, we introduce
the model setup and study the dynamics of the quantized EM field induced
by the magnetic dipoles. In Sec.\,\ref{sec3}, we derive the general
result for the field-induced interaction between two magnetic dipoles
with multiple states. In Sec.\,\ref{sec:TLS}, in the example of
two-level dipoles, we discuss the different interaction types according
to our general result. Sec.\,\ref{summary} is a brief summary.

\section{Field dynamics influenced by dipoles \label{sec2}}

We consider there are two magnetic dipoles placed in the EM field
($\hat{H}_{\text{\textsc{em}}}=\sum_{\mathbf{k}\sigma}\hbar\omega_{\mathbf{k}}\hat{a}_{\mathbf{k}\sigma}^{\dagger}\hat{a}_{\mathbf{k}\sigma}$).
Each dipole has multiple energy levels, and the self-Hamiltonian reads
$\hat{H}_{i}=\sum_{x}\text{\textsc{e}}_{i}^{(x)}|x\rangle_{i}\langle x|$
($i=1,2$ for dipole-1,2), where $\text{\textsc{e}}_{i}^{(x)}$ and
$|x\rangle_{i}$ are the eigen energy and the corresponding eigenstate. 

The two magnetic dipoles interact with the EM field via the interaction
\citep{wang2018magnetic,weinberg2012lectures} 
\begin{equation}
\hat{H}_{{\rm int}}=-\sum_{i=1,2}\hat{\mathfrak{m}}_{i}(t)\cdot\hat{\mathbf{B}}(\mathbf{x}_{i},t).\label{Eq.(1)}
\end{equation}
Here $\hat{\mathfrak{m}}_{i}$ is the magnetic dipole moment, $\mathbf{x}_{i}$
is the position of dipole-$i$. Under the Coulomb gauge, the quantized
magnetic field $\hat{\mathbf{B}}\left(\mathbf{x},t\right)=\nabla\times\hat{\mathbf{A}}\left(\mathbf{x},t\right)$
reads as 
\begin{equation}
\hat{\mathbf{B}}(\mathbf{x},t)=\sum_{\mathbf{k}\sigma}\mathrm{i}\hat{\mathrm{e}}_{\mathbf{k}\check{\sigma}}Z_{\mathbf{k}}\Big[\hat{a}_{\mathbf{k}\sigma}(t)e^{\mathrm{i}\mathbf{k}\cdot\mathbf{x}}-{\rm H.c.}\Big],\label{eq:B(x,t)}
\end{equation}
where $Z_{\mathbf{k}}:=\sqrt{\mu_{0}\hbar\omega_{\mathbf{k}}/2V}$
and $\hat{\mathrm{e}}_{\mathbf{k}\check{\sigma}}:=\hat{\mathrm{e}}_{\mathbf{k}}\times\hat{\mathrm{e}}_{\mathbf{k}\sigma}$.
The index $\check{\sigma}$ refers to the polarization direction $\hat{\mathrm{e}}_{\mathbf{k}\check{\sigma}}$
orthogonal to $\hat{\mathrm{e}}_{\mathbf{k}\sigma}$.

Generally, the dipole moment operator $\hat{\mathfrak{m}}_{i}$ always
can be expanded as $\hat{\mathfrak{m}}_{i}=\sum_{x,y}\vec{m}_{i}^{xy}\hat{\tau}_{i}^{xy}$,
with $\vec{m}_{i}^{xy}:=\langle x|\hat{\mathfrak{m}}_{i}|y\rangle_{i}$
as the dipole moment amplitude, and $\hat{\tau}_{i}^{xy}:=|x\rangle_{i}\langle y|$
as the unitless transition operator. 

The diagonal elements $\vec{m}_{i}^{xx}=\langle x|\hat{\mathfrak{m}}_{i}|x\rangle_{i}$
indicate the expectation value of the dipole moment on the level $|x\rangle_{i}$,
thus they are \emph{permanent dipoles}. The off-diagonal elements
$\vec{m}_{i}^{xy}=\langle x|\hat{\mathfrak{m}}_{i}|y\rangle_{i}$
($x\neq y$) are usually related to the photon radiation process,
and they are \emph{transition dipoles} \citep{wang2018magnetic}.

When the magnetic dipoles are placed in the EM field, they change
the original field dynamics. Then such influence propagates to each
other, and that is how the interaction is generated between the two
dipoles. 

To make further calculation, we first consider that only dipole-1
is placed in the vacuum field, and study the field dynamics induced
by dipole-1. With the above notations, such interaction $\hat{H}_{{\rm int}}^{(1)}=-\hat{\mathfrak{m}}_{1}(t)\cdot\hat{\mathbf{B}}(\mathbf{x}_{1},t)$
can be rewritten as 
\begin{align}
\hat{H}_{{\rm int}}^{(1)} & =\sum_{xy,\mathbf{k}\sigma}g_{1,\mathbf{k}\sigma}^{xy}\,\hat{\tau}_{1}^{xy}(t)\,\hat{a}_{\mathbf{k}\sigma}(t)+{\rm H.c.},\nonumber \\
g_{i,\mathbf{k}\sigma}^{xy} & =-\mathrm{i}\big(\vec{m}_{i}^{xy}\cdot\hat{\mathrm{e}}_{\mathbf{k}\check{\sigma}}\big)Z_{\mathbf{k}}e^{\mathrm{i}\mathbf{k}\cdot\mathbf{x}_{i}}.
\end{align}
 Here the coefficients $g_{i,\mathbf{k}\sigma}^{xy}$ enclose the
contributions from the EM field, the dipole moment $\vec{m}_{i}^{xy}$,
and the position $\mathbf{x}_{i}$.

Under this interaction, the field dynamics is given by the Heisenberg
equation $\frac{d}{dt}\hat{a}_{\mathbf{k}\sigma}=-\mathrm{i}\,\omega_{\mathbf{k}}\hat{a}_{\mathbf{k}\sigma}-\frac{\mathrm{i}}{\hbar}[\hat{a}_{\mathbf{k}\sigma},\hat{H}_{{\rm int}}^{(1)}]$,
which gives 
\[
\hat{a}_{\mathbf{k}\sigma}(t)=\hat{a}_{\mathbf{k}\sigma}(0)e^{\text{--}\mathrm{i}\omega_{\mathbf{k}}t}-\sum_{xy}\frac{\mathrm{i}}{\hbar}\big(g_{1,\mathbf{k}\sigma}^{xy}\big)^{*}\int_{0}^{t}ds\,e^{\text{--}\mathrm{i}\omega_{\mathbf{k}}s}\hat{\tau}_{1}^{yx}(t-s).
\]
Here the first term comes from the free field, and the second term
is induced by the interaction with dipole-1.

Now we put this $\hat{a}_{\mathbf{k}\sigma}(t)$ back into the above
magnetic field $\hat{\mathbf{B}}(\mathbf{x},t)$ {[}Eq.\,(\ref{eq:B(x,t)}){]},
and it can be rewritten as $\hat{\mathbf{B}}(\mathbf{x},t)=\hat{\mathbf{B}}_{0}(\mathbf{x},t)+\hat{\mathbf{B}}_{\text{D1}}(\mathbf{x},t)$,
where 
\begin{align}
\hat{\mathbf{B}}_{0}= & \sum_{\mathbf{k}\sigma}\mathrm{i}\hat{\mathrm{e}}_{\mathbf{k}\check{\sigma}}Z_{\mathbf{k}}\Big[\hat{a}_{\mathbf{k}\sigma}(0)e^{\mathrm{i}\mathbf{k}\cdot\mathbf{x}-\mathrm{i}\omega_{\mathbf{k}}t}-{\rm H.c.}\Big],\nonumber \\
\hat{\mathbf{B}}_{\text{D1}}= & \sum_{\mathbf{k}\sigma,xy}\frac{\hat{\mathrm{e}}_{\mathbf{k}\check{\sigma}}Z_{\mathbf{k}}}{\hbar}\,\big(g_{1,\mathbf{k}\sigma}^{xy}\big)^{*}e^{\mathrm{i}\mathbf{k}\cdot\mathbf{x}}\int_{0}^{t}ds\,e^{-\mathrm{i}\omega_{\mathbf{k}}s}\hat{\tau}_{1}^{yx}(t-s)\nonumber \\
 & +{\rm H.c.}\label{eq:BD-1}
\end{align}
Here $\hat{\mathbf{B}}_{0}(\mathbf{x},t)$ is the original free field
when there are no dipoles. The presence of dipole-1 changes the EM
field, and gives rise to  $\hat{\mathbf{B}}_{\text{D1}}(\mathbf{x},t)$,
which we call as the dipole field. When this dipole field propagates
to dipole-2, the dipole-dipole interaction is induced. 

\section{Interaction propagated by the dipole field \label{sec3} }

Now we consider dipole-2 is placed in the field ($\hat{\mathfrak{m}}_{2}=\sum_{uv}\vec{m}_{2}^{uv}\hat{\tau}_{2}^{uv}$),
which interacts with EM field locally via $\hat{H}_{{\rm int}}^{(2)}=-\hat{\mathfrak{m}}_{2}(t)\cdot\hat{\mathbf{B}}(\mathbf{x}_{2},t)$.
Clearly, the dipole-dipole interaction is the part induced by the
above dipole field $\hat{\mathbf{B}}_{\text{D1}}(\mathbf{x},t)$ {[}Eq.\,(\ref{eq:BD-1}){]},
which can be rewritten as \begin{widetext}
\begin{align}
\hat{H}_{1\rightarrow2}= & -\hat{\mathfrak{m}}_{2}(t)\cdot\hat{\mathbf{B}}_{\text{D1}}(\mathbf{x}_{2},t)=\sum_{uv}\sum_{\mathbf{k}\sigma,xy}-\frac{\mathrm{i}}{\hbar}\big(g_{1,\mathbf{k}\sigma}^{xy}\big)^{*}g_{2,\mathbf{k}\sigma}^{uv}\int_{0}^{t}ds\,e^{-\mathrm{i}\omega_{\mathbf{k}}s}\hat{\tau}_{1}^{yx}(t-s)\cdot\hat{\tau}_{2}^{uv}(t)+{\rm H.c.}\nonumber \\
= & \sum_{uv,xy}\int_{0}^{t}ds\,\mathcal{D}_{1\rightarrow2}^{yx,uv}(s)\,\hat{\tau}_{1}^{yx}(t-s)\cdot\hat{\tau}_{2}^{uv}(t)+{\rm H.c.}\label{eq:H1to2-Int}
\end{align}
 In the last line, we collect the mode summation of $\mathbf{k}\sigma$
into a convolution kernel $\mathcal{D}_{1\rightarrow2}^{yx,uv}(s)$,
which is defined as 
\begin{equation}
\mathcal{D}_{1\rightarrow2}^{yx,uv}(s):=\sum_{\mathbf{k}\sigma}-\frac{\mathrm{i}}{\hbar}\big(g_{1,\mathbf{k}\sigma}^{xy}\big)^{*}g_{2,\mathbf{k}\sigma}^{uv}\,e^{-\mathrm{i}\omega_{\mathbf{k}}s}:=-\mathrm{i}\int_{0}^{\infty}\frac{d\omega}{2\pi}\,J_{1\rightarrow2}^{yx,uv}(\omega)e^{-\mathrm{i}\omega s}.
\end{equation}
Here we define $J_{1\rightarrow2}^{yx,uv}(\omega):=\frac{2\pi}{\hbar}\sum_{\mathbf{k}\sigma}\big(g_{1,\mathbf{k}\sigma}^{xy}\big)^{*}g_{2,\mathbf{k}\sigma}^{uv}\delta(\omega-\omega_{\mathbf{k}})$
as the \emph{coupling spectral density}, which can be used to turn
the above summation of \textbf{$\mathbf{k}\sigma$ }into a continuous
integral \citep{breuer_theory_2002,wang2018magnetic}. For the free-space
EM field it can be explicitly calculated out as (see Appendix \ref{sec:Coupling-spectral-density})
\begin{align}
J_{1\rightarrow2}^{yx,uv}(\omega)= & \frac{\mu_{0}}{2\pi r^{3}}\Big\{\vec{m}_{1}^{yx}\cdot\vec{m}_{2}^{uv}\big[\eta^{2}\sin\eta+\eta\cos\eta-\sin\eta\big]\nonumber \\
 & -3(\vec{m}_{1}^{yx}\cdot\hat{\mathrm{e}}_{\mathbf{r}})(\vec{m}_{2}^{uv}\cdot\hat{\mathrm{e}}_{\mathbf{r}})\,\big[\frac{1}{3}\eta^{2}\sin\eta+\eta\cos\eta-\sin\eta\big]\Big\},\label{eq:J12}
\end{align}
where we denote $\eta:=kr=\omega r/c$, $c$ is the velocity of light,
and $r=|\mathbf{x}_{1}-\mathbf{x}_{2}|$ is the distance between the
two dipoles. We can see $J_{1\rightarrow2}^{yx,uv}(\omega)=-J_{1\rightarrow2}^{yx,uv}(-\omega)$
is an odd function, and $J_{2\rightarrow1}^{uv,yx}(\omega)=J_{1\rightarrow2}^{yx,uv}(\omega)$.
When $\omega\rightarrow0$, we have $J_{1\rightarrow2}^{yx,uv}(\omega)\rightarrow0$.
\end{widetext}

The interaction (\ref{eq:H1to2-Int}) naturally has a retarded form,
and up to now it is still an exact result without any approximation.
Here the kernel function $\mathcal{D}_{1\rightarrow2}^{yx,uv}(s)$
characterizes the propagating process of the interaction, and it is
a fasting-decaying function of $t$, because its (half-)Fourier image
$J_{1\rightarrow2}^{yx,uv}(\omega)$ has a quite wide spectrum.

Since $\mathcal{D}_{1\rightarrow2}^{yx,uv}(t)$ is already of the
second order of the field-dipole interaction $g_{i,\mathbf{k}\sigma}^{xy,uv}$,
thus we make an approximation \citep{lehmberg1970radiation,wang2018magnetic}
\begin{equation}
\hat{\tau}_{1}^{yx}(t-s)\simeq\hat{\tau}_{1}^{yx}(t)\cdot e^{\mathrm{i}\Omega_{1}^{yx}s},
\end{equation}
where $\Omega_{1}^{yx}:=(\text{\textsc{e}}_{1}^{(y)}-\text{\textsc{e}}_{1}^{(x)})/\hbar$
is the transition frequency between the two energy levels $|y\rangle_{1}$
and $|x\rangle_{1}$. That means, its dynamics is governed only by
the self-Hamiltonian $\hat{H}_{1}$, without considering its interaction
with the field of higher orders. Then the interaction (\ref{eq:H1to2-Int})
would become local in time, which is composed of summation terms $\sim\hat{\tau}_{1}^{yx}(t)\cdot\hat{\tau}_{2}^{uv}(t)$
multiplying by time-independent interaction strengths.

Further, since $\mathcal{D}_{1\rightarrow2}^{yx,uv}(t)$ decays very
fast to zero, we can extend the time integral in Eq.\,(\ref{eq:H1to2-Int})
to $t\rightarrow\infty$, and that reduces the interaction Hamiltonian
as
\begin{equation}
\hat{H}_{1\rightarrow2}=\sum_{xy,uv}\xi_{1\rightarrow2}^{yx,uv}\,\hat{\tau}_{1}^{yx}\hat{\tau}_{2}^{uv}+\mathrm{H.c.},\label{eq:H1to2}
\end{equation}
where $\xi_{1\rightarrow2}^{yx,uv}$ is calculated by\footnote{Here
we used the relation $\int_{0}^{\infty}ds\,e^{\mathrm{i}\omega s}=\pi\delta(\omega)+\mathrm{i}\mathbf{P}\frac{1}{\omega}$,
and $\mathbf{P}$ means the principal integral.} 
\begin{align}
 & \xi_{1\rightarrow2}^{yx,uv}=-\mathrm{i}\int_{0}^{t\rightarrow\infty}ds\int_{0}^{\infty}\frac{d\omega}{2\pi}\,J_{1\rightarrow2}^{yx,uv}(\omega)e^{\mathrm{i}(\Omega_{1}^{yx}-\omega)s}\nonumber \\
= & \mathbf{P}\int_{0}^{\infty}\frac{d\omega}{2\pi}\,\frac{J_{1\rightarrow2}^{yx,uv}(\omega)}{\Omega_{1}^{yx}-\omega}-\frac{\mathrm{i}}{2}\int_{0}^{\infty}d\omega J_{1\rightarrow2}^{yx,uv}(\omega)\delta(\Omega_{1}^{yx}-\omega).
\end{align}

By rearranging the summation indices, the interaction (\ref{eq:H1to2-Int})
can be rewritten as 
\begin{align}
\hat{H}_{1\rightarrow2} & =\sum_{xy,uv}\varLambda_{1\rightarrow2}^{yx,uv}\,\hat{\tau}_{1}^{yx}\hat{\tau}_{2}^{uv},\nonumber \\
\varLambda_{1\rightarrow2}^{yx,uv} & =\xi_{1\rightarrow2}^{yx,uv}+(\xi_{1\rightarrow2}^{xy,vu})^{*}\nonumber \\
 & =K_{1\rightarrow2}^{yx,uv}\big(\Omega_{1}^{yx}\big)-\frac{\mathrm{i}}{2}J_{1\rightarrow2}^{yx,uv}\big(\Omega_{1}^{yx}\big),
\end{align}
where $K_{1\rightarrow2}^{yx,uv}\big(\Omega_{1}^{yx}\big)$ is given
by the function
\begin{align}
 & K_{1\rightarrow2}^{yx,uv}(\Omega):=\mathbf{P}\int_{-\infty}^{\infty}\frac{d\omega}{2\pi}\,\frac{J_{1\rightarrow2}^{yx,uv}(\omega)}{\Omega-\omega}\nonumber \\
= & \frac{\mu_{0}}{4\pi r^{3}}\Big\{\vec{m}_{1}^{yx}\cdot\vec{m}_{2}^{uv}[\cos\eta-\eta\sin\eta-\eta^{2}\cos\eta]\nonumber \\
- & 3(\vec{m}_{1}^{yx}\cdot\hat{\mathrm{e}}_{\mathbf{r}})(\vec{m}_{2}^{uv}\cdot\hat{\mathrm{e}}_{\mathbf{r}})[\cos\eta-\eta\sin\eta-\frac{1}{3}\eta^{2}\cos\eta]\Big\},\label{eq:K12}
\end{align}
with the ratio $\eta=\Omega r/c$. Here we have utilized the relations
$\Omega_{1}^{yx}=-\Omega_{1}^{xy}$, $J_{1\rightarrow2}^{yx,uv}(-\omega)=-J_{1\rightarrow2}^{yx,uv}(\omega)$,
and $[J_{1\rightarrow2}^{xy,vu}(\omega)]^{*}=J_{1\rightarrow2}^{yx,uv}(\omega)$.
Notice that $K_{1\rightarrow2}^{yx,uv}(\Omega)=K_{1\rightarrow2}^{yx,uv}(-\Omega)$
is an even function. 

Following the same method as above, we should also consider the symmetric
procedure that dipole-2 generates a dipole field $\hat{\mathbf{B}}_{\text{D2}}(\mathbf{x},t)$
and then interacts with dipole-1. That results to an interaction Hamiltonian
$\hat{H}_{2\rightarrow1}=\sum\varLambda_{2\rightarrow1}^{uv,yx}\,\hat{\tau}_{2}^{uv}\hat{\tau}_{1}^{yx}$,
which can be simply obtained by making proper variable exchanges from
the above results. Finally, the total interaction between the two
dipoles should be the average of the these two symmetric procedures
$\hat{H}_{1\leftrightarrow2}=\frac{1}{2}(\hat{H}_{1\rightarrow2}+\hat{H}_{2\rightarrow1})$,
and we rewrite it as
\begin{align}
\hat{H}_{1\leftrightarrow2} & :=\sum_{xy,uv}(G_{yx,uv}^{(\mathtt{P})}+G_{yx,uv}^{(\mathtt{D})})\hat{\tau}_{1}^{yx}\hat{\tau}_{2}^{uv},\nonumber \\
G_{yx,uv}^{(\mathtt{P})} & =\frac{1}{2}[K_{1\rightarrow2}^{yx,uv}(\Omega_{1}^{yx})+K_{2\rightarrow1}^{uv,yx}(\Omega_{2}^{uv})],\label{eq:Res}\\
G_{yx,uv}^{(\mathtt{D})} & =\frac{1}{4\mathrm{i}}\big[J_{1\rightarrow2}^{yx,uv}(\Omega_{1}^{yx})+J_{2\rightarrow1}^{uv,yx}(\Omega_{2}^{uv})\big].\nonumber 
\end{align}
Here the interaction strength is divided into two parts: the \emph{principal
term} $G_{yx,uv}^{(\mathtt{P})}$ comes from the principal integral
which is associated with virtual photon exchange, while the \emph{dissipation
correction} $G_{yx,uv}^{(\mathtt{D})}$ is associated with the real
photon emission in the atom decay process \citep{li2015steady,li2016non}.
Notice here $J_{i\rightarrow j}^{yx,uv}(\cdot)$ and $K_{i\rightarrow j}^{yx,uv}(\cdot)$
are not necessarily real numbers, which are determined by the vectors
$\vec{m}_{i}^{yx}=\langle y|\hat{\mathfrak{m}}_{i}|x\rangle_{i}$. 

Now we derived the general result for the field-induced interaction
between two magnetic dipoles. No RWA was adopted, and there was no
special requirement for the transition frequencies $\Omega_{i}^{yx}$.
In the following, we will clarify the physical meaning of the different
interaction terms in this result with the help of an example on a
pair of two-level dipoles.

\section{A pair of two-level dipoles \label{sec:TLS}}

\begin{figure}
\includegraphics[width=0.25\textwidth]{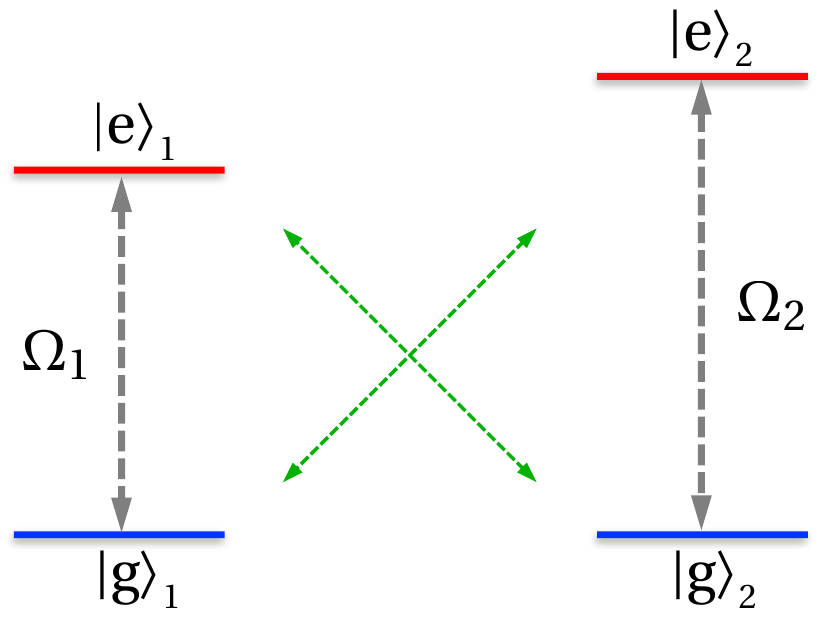}

\caption{Demonstration for a pair of two-level dipoles. Here $\Omega_{i}:=\Omega_{i}^{\text{eg}}=(\text{\textsc{e}}_{i}^{(\mathrm{e})}-\text{\textsc{e}}_{i}^{(\mathrm{g})})/\hbar$
is the transition frequency for $i=1,2$.}

\label{fig-TLS}
\end{figure}
The above result (\ref{eq:Res}) applies for general multi-state systems,
and contains the interaction between any two matrix elements of the
two dipole operators $\hat{\mathfrak{m}}_{i}=\sum\vec{m}_{i}^{xy}\hat{\tau}_{i}^{xy}$.
Now we focus on an example that the two dipoles both have only two
energy levels (Fig.\,\ref{fig-TLS}), and discuss the different interaction
types in the above general result (\ref{eq:Res}) separately. 

\vspace{1em}\noindent \textbf{Permanent dipoles: }

We first look at the interaction between the diagonal terms 
\begin{equation}
\hat{H}_{xx,yy}:=(G_{xx,yy}^{(\mathtt{P})}+G_{xx,yy}^{(\mathtt{D})})\hat{\tau}_{1}^{xx}\hat{\tau}_{2}^{yy},
\end{equation}
 where $\hat{\tau}_{i}^{xx}=|x\rangle_{i}\langle x|$ for $x=\mathrm{e},\mathrm{g}$.
In the interaction strength $G_{xx,yy}^{(...)}$, the dipole moment
vector $\vec{m}_{i}^{xx}=\langle x|\hat{\mathfrak{m}}_{i}|x\rangle_{i}$
is the expectation value of the dipole operator $\hat{\mathfrak{m}}_{i}$
on the state $|x\rangle_{i}$, thus we say $\hat{\mathfrak{m}}_{i}^{xx}:=\vec{m}_{i}^{xx}\hat{\tau}_{i}^{xx}$
is the permanent dipole operator.

For these permanent dipole moments, the transition frequencies are
$\Omega_{i}^{xx}=0$, thus the dissipation correction vanishes $G_{xx,yy}^{(\mathtt{D})}=0$,
and the principal term $G_{xx,yy}^{(\mathtt{P})}$ directly returns
the classical result {[}see $\Omega\rightarrow0$ in Eqs.\,(\ref{eq:K12},
\ref{eq:Res}){]}. Finally, such interaction terms can be written
as 
\begin{equation}
\hat{H}_{xx,yy}=\frac{\mu_{0}}{4\pi r^{3}}[\hat{\mathfrak{m}}_{1}^{xx}\cdot\hat{\mathfrak{m}}_{2}^{yy}-3(\hat{\mathfrak{m}}_{1}^{xx}\cdot\hat{\mathrm{e}}_{\mathbf{r}})(\hat{\mathfrak{m}}_{1}^{yy}\cdot\hat{\mathrm{e}}_{\mathbf{r}})],
\end{equation}
which has exactly the same form with the classical one.

\vspace{1em}\noindent \textbf{Resonant transition dipoles: }

Now we consider the interaction term between resonant transition dipoles.
For the transition operators $\hat{\tau}_{1}^{+}:=|\mathrm{e}\rangle_{_{1}}\langle\mathrm{g}|$,
$\hat{\tau}_{2}^{-}:=|\mathrm{g}\rangle_{_{2}}\langle\mathrm{e}|$,
with resonant transition frequencies $\Omega_{1}^{\text{eg}}=-\Omega_{2}^{\text{ge}}:=\Omega$,
the corresponding interaction term is 
\begin{equation}
\hat{H}_{\text{eg,ge}}=(G_{\text{eg,ge}}^{(\mathtt{P})}+G_{\text{eg,ge}}^{(\mathtt{D})})\,\hat{\tau}_{1}^{+}\hat{\tau}_{2}^{-}.
\end{equation}

From Eq.\,(\ref{eq:Res}) we can see the two summation terms in $G_{\text{eg,ge}}^{(\mathtt{D})}$
exactly cancel each other {[}since $J_{2\rightarrow1}^{uv,yx}(\omega)=J_{1\rightarrow2}^{yx,uv}(\omega)${]},
thus the dissipation term always vanishes ($G_{\text{eg,ge}}^{(\mathtt{D})}=0$).
The two terms in the principal term equal to each other, and that
gives $G_{\text{eg,ge}}^{(\mathtt{P})}=K_{1\rightarrow2}^{\text{eg,ge}}(\Omega)$.
This result is the same as previous calculations based on master equations
\citep{lehmberg1970radiation,agarwal_quantum_1974,ficek1987quantum,ficek2005quantum,wang2018magnetic}.
This principal interaction strength does not return to the classical
form directly, but returns at the Dicke limit $r/\lambda\rightarrow0$,
where $\lambda$ is the wavelength of the transition frequency $\Omega$.

Following the same discussion, there is also a conjugate term for
the transition dipole interaction $\hat{H}_{\text{ge,eg}}=G_{\text{ge,eg}}^{(\mathtt{P})}\,\hat{\tau}_{1}^{-}\hat{\tau}_{2}^{+}$,
with $G_{\text{ge,eg}}^{(\mathtt{P})}=[G_{\text{eg,ge}}^{(\mathtt{P})}]^{*}$.

\vspace{1em}\noindent \textbf{Non-resonant transition dipoles:}

Further, we consider the case that frequencies of the two transition
dipoles are different $\Omega_{1}\neq\Omega_{2}$. In this case, the
above result (\ref{eq:Res}) gives the principal term and the dissipation
correction as 
\begin{align}
G_{\text{eg,ge}}^{(\mathtt{P})} & =\frac{1}{2}\big[K_{1\rightarrow2}^{\text{eg,ge}}(\Omega_{1})+K_{1\rightarrow2}^{\text{eg,ge}}(\Omega_{2})\big],\nonumber \\
G_{\text{eg,ge}}^{(\mathtt{D})} & =\frac{1}{4\mathrm{i}}\big[J_{1\rightarrow2}^{\text{eg,ge}}(\Omega_{1})-J_{1\rightarrow2}^{\text{eg,ge}}(\Omega_{2})\big].\label{eq:non-resonant}
\end{align}
Since $\Omega_{1}\neq\Omega_{2}$, the two terms in $G_{\text{eg,ge}}^{(\mathtt{D})}$
can no longer cancel each other completely. Notice that, when deriving
the GKSL (Lindblad) master equation for an atomic system with coherent
transitions, there is also such a similar unitary correction term
like $G_{\text{eg,ge}}^{(\mathtt{D})}$ (see Eq.\,(12) in Ref.\,\citep{li2015steady}
or Eq.\,(17) in Ref.\,\citep{li2016non}), which comes from the
dissipation terms of the master equation. Therefore, here we say $G_{xy,uv}^{(\mathtt{D})}$
is the dissipation correction. Similarly, the principal term becomes
the average of $K_{1\rightarrow2}^{\text{eg,ge}}(\Omega_{1})$ and
$K_{1\rightarrow2}^{\text{eg,ge}}(\Omega_{2})$.

From this result we can see the two dipoles emit both real and virtual
photons, and they have interference with each other. But due to the
frequency difference, they cannot achieve the complete interference
as the above resonance case, thus the two terms in $G_{\text{eg,ge}}^{(\mathtt{D})}$
cannot cancel each other exactly.

\vspace{1em}\noindent \textbf{Counter-rotating terms and mixed terms: }

It is worth noticing that the above interaction (\ref{eq:Res}) also
contains counter-rotating terms, such as 
\begin{align}
\hat{H}_{\text{eg,eg}} & =(G_{\text{eg,eg}}^{(\mathtt{P})}+G_{\text{eg,eg}}^{(\mathtt{D})})\,\hat{\tau}_{1}^{+}\hat{\tau}_{2}^{+},\nonumber \\
G_{\text{eg,eg}}^{(\mathtt{P})} & =\frac{1}{2}\big[K_{1\rightarrow2}^{\text{eg,eg}}(\Omega_{1})+K_{1\rightarrow2}^{\text{eg,eg}}(\Omega_{2})\big],\nonumber \\
G_{\text{eg,eg}}^{(\mathtt{D})} & =\frac{1}{4\mathrm{i}}\big[J_{1\rightarrow2}^{\text{eg,eg}}(\Omega_{1})+J_{1\rightarrow2}^{\text{eg,eg}}(\Omega_{2})\big].
\end{align}
 Here the principal interaction strength $G_{\text{eg,eg}}^{(\mathtt{P})}$
is quite similar with the above transition interaction (\ref{eq:non-resonant})
(except the ordering difference in the superscript). The dissipation
term $G_{\text{eg,eg}}^{(\mathtt{D})}$ always exists for nonzero
$\Omega_{i}$.

Moreover, indeed the interaction (\ref{eq:Res}) also contains ``mixed''
interaction terms between the transition and permanent dipoles, e.g.,
\begin{align}
\hat{H}_{\text{eg,gg}} & =(G_{\text{eg,gg}}^{(\mathtt{P})}+G_{\text{eg,gg}}^{(\mathtt{D})})\,\hat{\tau}_{1}^{+}\hat{\tau}_{2}^{\text{gg}},\nonumber \\
G_{\text{eg,gg}}^{(\mathtt{P})} & =\frac{1}{2}\big[K_{1\rightarrow2}^{\text{eg,gg}}(\Omega_{1})+K_{1\rightarrow2}^{\text{eg,gg}}(0)\big]\nonumber \\
G_{\text{eg,gg}}^{(\mathtt{D})} & =\frac{1}{4\mathrm{i}}J_{1\rightarrow2}^{\text{eg,gg}}(\Omega_{1}),
\end{align}
 where $K_{1\rightarrow2}^{\text{eg,gg}}(0)$ has the classical interaction
form.

Usually, these counter-rotating and mixed interaction terms do not
exhibit significant physical effects since they oscillate too fast
($\sim e^{\pm\mathrm{i}(\Omega_{1}+\Omega_{2})t}$, or $\sim e^{\pm\mathrm{i}\Omega_{i}t}$)
comparing with the typical time scale determined by the interaction
energy $G_{xy,uv}^{(...)}$, thus averagely they do not have significant
contributions and  can be omitted by the RWA. When the detuning between
the two frequencies is quite large (e.g., $\Omega_{1}\gg\Omega_{2}$),
we have $\Omega_{1}-\Omega_{2}\simeq\Omega_{1}+\Omega_{2}\simeq\Omega_{1}$,
then these oscillating terms could be comparable with the non-resonant
transition terms.

From this example of two-level dipoles, we can see the above general
result (\ref{eq:Res}) indeed contains different types of magnetic
dipole-dipole interactions, and they are derived altogether based
on the field propagation. Clearly, the generalization to multi-state
systems is straightforward. 

In the Dicke limit ($\Omega_{1,2}r/c\rightarrow0$), we can see all
the dissipation corrections approach zero, and all the principal terms
approach $K_{1\rightarrow2}^{yx,uv}(0)$. Thus, after summing up all
the above terms, the interaction can be written as
\begin{align}
\hat{H} & =\sum_{xy,uv}K_{1\rightarrow2}^{yx,uv}(0)\,\hat{\tau}_{1}^{yx}\hat{\tau}_{2}^{uv}\nonumber \\
 & =\sum_{xy,uv}\frac{\mu_{0}}{4\pi r^{3}}\big[\vec{m}_{1}^{yx}\cdot\vec{m}_{2}^{uv}-3(\vec{m}_{1}^{yx}\cdot\hat{\mathrm{e}}_{\mathbf{r}})(\vec{m}_{2}^{uv}\cdot\hat{\mathrm{e}}_{\mathbf{r}})\big]\hat{\tau}_{1}^{yx}\hat{\tau}_{2}^{uv}\nonumber \\
 & =\frac{\mu_{0}}{4\pi r^{3}}\big[\hat{\mathfrak{m}}_{1}\cdot\hat{\mathfrak{m}}_{2}-3(\hat{\mathfrak{m}}_{1}\cdot\hat{\mathrm{e}}_{\mathbf{r}})(\hat{\mathfrak{m}}_{2}\cdot\hat{\mathrm{e}}_{\mathbf{r}})\big],
\end{align}
which just returns the classical form (\ref{eq:classical}). In this
sense, such field-induced interaction is consistent with that directly
exchanged from the classical interaction. If the distance between
the two dipoles is not short enough, the quantum corrections in the
off-diagonal interaction terms (the transition terms, counter-rotating
terms, and the mixed terms) should be considered.

\section{Summary \label{summary}}

In this paper we made a general derivation for the magnetic dipole-dipole
interaction based on the propagation of the quantized EM field. In
this derivation, the two dipoles may have multiple energy levels,
and we only adopted the second-order perturbation and the Markovian
approximation, but did not require the resonance condition. Since
the RWA is no longer needed, our result naturally gives the DD interaction
for non-resonant dipoles, and even the counter-rotating interaction
terms, as well as the mixed interaction terms between the permanent
and transition dipoles. As long as the dipole operator $\hat{\mathfrak{m}}=\sum\vec{m}_{xy}|x\rangle\langle y|$
is written down for a specific physical system, all these interaction
terms can be derived based on the field mediation.

From this result, we can notice that this field-induced DD interaction
is associated with the inference of the virtual/real photons emitted
from the two dipoles. As a result, the interaction strength could
be influenced by the frequency difference of the two dipoles. If the
two dipole moments has the same frequency, our result naturally returns
to the result in previous studies. In principle, our derivation may
also apply if the EM field has special properties (e.g., by applying
certain periodic modulation or placing the system in a cavity with
finite size \citep{wang2018magnetic,baranov_magnetic_2016,donaire_dipole-dipole_2017,tang_tuning_2018,davis_photon-mediated_2019}),
or if the mediation field is not the EM field (e.g., the phonon field). 

\vspace{0.8em}

\emph{Acknowledgements }- This study is supported by the National
Natural Science Foundation of China (Grant No. 11675014), the Ministry
of Science and Technology of China (2013YQ030595-3), and the BIT Research
Fund Program for Young Scholars. S.-W. Li appreciated quite much for
the helpful discussions with N. Wu and D. Xu in Beijing Institute
of Technology.

\appendix
\begin{widetext}

\section{The coupling spectral density in free space \label{sec:Coupling-spectral-density}}

Here we show the derivation of the coupling spectral density $J_{1\rightarrow2}^{yx,uv}(\omega)$
in free space, which is define by 
\begin{equation}
J_{1\rightarrow2}^{yx,uv}(\omega):=\frac{2\pi}{\hbar}\sum_{\mathbf{k}\sigma}\big(g_{1,\mathbf{k}\sigma}^{xy}\big)^{*}g_{2,\mathbf{k}\sigma}^{uv}\delta(\omega-\omega_{\mathbf{k}}),\qquad g_{i,\mathbf{k}\sigma}^{xy}=-\mathrm{i}\big(\vec{m}_{i}^{xy}\cdot\hat{\mathrm{e}}_{\mathbf{k}\check{\sigma}}\big)\sqrt{\frac{\mu_{0}\hbar\omega_{\mathbf{k}}}{2V}}e^{\mathrm{i}\mathbf{k}\cdot\mathbf{x}_{i}}.
\end{equation}
 The summation over $\mathbf{k},\sigma$ is changed into integral
by
\begin{equation}
\sum_{\mathbf{k}\sigma}[...]\quad\longrightarrow\quad\sum_{\sigma}\frac{V}{(2\pi)^{3}}\int d^{3}\mathbf{k}\,[...]=\sum_{\sigma}\frac{V}{(2\pi c)^{3}}\int\omega^{2}d\omega\int d\Omega\,[...]
\end{equation}
Thus the coupling spectral density $J_{1\rightarrow2}^{yx,uv}(\omega)$
is given by (denoting $\mathbf{r}:=\mathbf{x}_{2}-\mathbf{x}_{1}$)
\begin{align}
J_{1\rightarrow2}^{yx,uv}(\omega) & =\frac{2\pi}{\hbar}\cdot\frac{V\omega^{2}}{(2\pi c)^{3}}\cdot\frac{\mu_{0}\hbar\omega}{2V}\int_{0}^{2\pi}d\varphi\int_{0}^{\pi}\sin\theta d\theta\,e^{\mathrm{i}\mathbf{k}\cdot(\mathbf{x}_{2}-\mathbf{x}_{1})}\Big[\sum_{\sigma}(\vec{m}_{1}^{yx}\cdot\hat{\mathrm{e}}_{\mathbf{k}\check{\sigma}})\,(\vec{m}_{2}^{uv}\cdot\hat{\mathrm{e}}_{\mathbf{k}\check{\sigma}})\Big]\nonumber \\
 & =\frac{1}{(2\pi)^{2}c^{3}}\,\frac{\mu_{0}\omega^{3}}{2}\int_{0}^{2\pi}d\varphi\int_{0}^{\pi}\sin\theta d\theta\,e^{\mathrm{i}\mathbf{k}\cdot\mathbf{r}}\Big[\vec{m}_{1}^{yx}\cdot\vec{m}_{2}^{uv}-(\vec{m}_{1}^{yx}\cdot\hat{\mathrm{e}}_{\mathbf{k}})\,(\vec{m}_{2}^{uv}\cdot\hat{\mathrm{e}}_{\mathbf{k}})\Big].
\end{align}
 Finally this integral gives (denoting $k:=\omega/c$, see Ref.\,\citep{wang2018magnetic})
\begin{align}
J_{1\rightarrow2}^{yx,uv}(\omega)=\frac{\mu_{0}k^{3}}{2\pi}\Big\{ & \vec{m}_{1}^{yx}\cdot\vec{m}_{2}^{uv}\big[\frac{\sin kr}{kr}+\frac{\cos kr}{(kr)^{2}}-\frac{\sin kr}{(kr)^{3}}\big]\nonumber \\
 & -(\vec{m}_{1}^{yx}\cdot\hat{\mathrm{e}}_{\mathbf{r}})\,(\vec{m}_{2}^{uv}\cdot\hat{\mathrm{e}}_{\mathbf{r}})\,\big[\frac{\sin kr}{kr}+\frac{3\cos kr}{(kr)^{2}}-\frac{3\sin kr}{(kr)^{3}}\big]\Big\}.
\end{align}
From this expression, we can see $J_{1\rightarrow2}^{yx,uv}(-\omega)=-J_{1\rightarrow2}^{yx,uv}(\omega)$
is an odd function of $\omega$. And we also have the following relation
when exchanging the indices:
\begin{equation}
J_{2\rightarrow1}^{uv,yx}(\omega):=\frac{2\pi}{\hbar}\sum_{\mathbf{k}\sigma}\big(g_{2,\mathbf{k}\sigma}^{vu}\big)^{*}g_{1,\mathbf{k}\sigma}^{yx}\delta(\omega-\omega_{\mathbf{k}})=[J_{1\rightarrow2}^{xy,vu}(\omega)]^{*}=J_{1\rightarrow2}^{yx,uv}(\omega).
\end{equation}
 \end{widetext}

\bibliographystyle{apsrev4-1}
\bibliography{reference,Refs}

\begin{thebibliography}{32}%
\makeatletter
\providecommand \@ifxundefined [1]{%
 \@ifx{#1\undefined}
}%
\providecommand \@ifnum [1]{%
 \ifnum #1\expandafter \@firstoftwo
 \else \expandafter \@secondoftwo
 \fi
}%
\providecommand \@ifx [1]{%
 \ifx #1\expandafter \@firstoftwo
 \else \expandafter \@secondoftwo
 \fi
}%
\providecommand \natexlab [1]{#1}%
\providecommand \enquote  [1]{``#1''}%
\providecommand \bibnamefont  [1]{#1}%
\providecommand \bibfnamefont [1]{#1}%
\providecommand \citenamefont [1]{#1}%
\providecommand \href@noop [0]{\@secondoftwo}%
\providecommand \href [0]{\begingroup \@sanitize@url \@href}%
\providecommand \@href[1]{\@@startlink{#1}\@@href}%
\providecommand \@@href[1]{\endgroup#1\@@endlink}%
\providecommand \@sanitize@url [0]{\catcode `\\12\catcode `\$12\catcode
  `\&12\catcode `\#12\catcode `\^12\catcode `\_12\catcode `\%12\relax}%
\providecommand \@@startlink[1]{}%
\providecommand \@@endlink[0]{}%
\providecommand \url  [0]{\begingroup\@sanitize@url \@url }%
\providecommand \@url [1]{\endgroup\@href {#1}{\urlprefix }}%
\providecommand \urlprefix  [0]{URL }%
\providecommand \Eprint [0]{\href }%
\providecommand \doibase [0]{http://dx.doi.org/}%
\providecommand \selectlanguage [0]{\@gobble}%
\providecommand \bibinfo  [0]{\@secondoftwo}%
\providecommand \bibfield  [0]{\@secondoftwo}%
\providecommand \translation [1]{[#1]}%
\providecommand \BibitemOpen [0]{}%
\providecommand \bibitemStop [0]{}%
\providecommand \bibitemNoStop [0]{.\EOS\space}%
\providecommand \EOS [0]{\spacefactor3000\relax}%
\providecommand \BibitemShut  [1]{\csname bibitem#1\endcsname}%
\let\auto@bib@innerbib\@empty
%</preamble>
\bibitem [{\citenamefont {Oh}\ \emph {et~al.}(2008)\citenamefont {Oh},
  \citenamefont {Huang}, \citenamefont {Peskin},\ and\ \citenamefont
  {Kais}}]{oh2008entanglement}%
  \BibitemOpen
  \bibfield  {author} {\bibinfo {author} {\bibfnamefont {S.}~\bibnamefont
  {Oh}}, \bibinfo {author} {\bibfnamefont {Z.}~\bibnamefont {Huang}}, \bibinfo
  {author} {\bibfnamefont {U.}~\bibnamefont {Peskin}}, \ and\ \bibinfo {author}
  {\bibfnamefont {S.}~\bibnamefont {Kais}},\ }\href@noop {} {\bibfield
  {journal} {\bibinfo  {journal} {Physical Review A}\ }\textbf {\bibinfo
  {volume} {78}},\ \bibinfo {pages} {062106} (\bibinfo {year}
  {2008})}\BibitemShut {NoStop}%
\bibitem [{\citenamefont {Hu}\ \emph {et~al.}(2018)\citenamefont {Hu},
  \citenamefont {Jia},\ and\ \citenamefont {Zhao}}]{hu2018effect}%
  \BibitemOpen
  \bibfield  {author} {\bibinfo {author} {\bibfnamefont {F.}~\bibnamefont
  {Hu}}, \bibinfo {author} {\bibfnamefont {W.}~\bibnamefont {Jia}}, \ and\
  \bibinfo {author} {\bibfnamefont {Q.}~\bibnamefont {Zhao}},\ }\href@noop {}
  {\bibfield  {journal} {\bibinfo  {journal} {Annals of Physics}\ }\textbf
  {\bibinfo {volume} {392}},\ \bibinfo {pages} {1} (\bibinfo {year}
  {2018})}\BibitemShut {NoStop}%
\bibitem [{\citenamefont {Balasubramanian}\ \emph {et~al.}(2008)\citenamefont
  {Balasubramanian}, \citenamefont {Chan}, \citenamefont {Kolesov},
  \citenamefont {Al-Hmoud}, \citenamefont {Tisler}, \citenamefont {Shin},
  \citenamefont {Kim}, \citenamefont {Wojcik}, \citenamefont {Hemmer},
  \citenamefont {Krueger} \emph {et~al.}}]{balasubramanian2008nanoscale}%
  \BibitemOpen
  \bibfield  {author} {\bibinfo {author} {\bibfnamefont {G.}~\bibnamefont
  {Balasubramanian}}, \bibinfo {author} {\bibfnamefont {I.}~\bibnamefont
  {Chan}}, \bibinfo {author} {\bibfnamefont {R.}~\bibnamefont {Kolesov}},
  \bibinfo {author} {\bibfnamefont {M.}~\bibnamefont {Al-Hmoud}}, \bibinfo
  {author} {\bibfnamefont {J.}~\bibnamefont {Tisler}}, \bibinfo {author}
  {\bibfnamefont {C.}~\bibnamefont {Shin}}, \bibinfo {author} {\bibfnamefont
  {C.}~\bibnamefont {Kim}}, \bibinfo {author} {\bibfnamefont {A.}~\bibnamefont
  {Wojcik}}, \bibinfo {author} {\bibfnamefont {P.~R.}\ \bibnamefont {Hemmer}},
  \bibinfo {author} {\bibfnamefont {A.}~\bibnamefont {Krueger}},  \emph
  {et~al.},\ }\href@noop {} {\bibfield  {journal} {\bibinfo  {journal}
  {Nature}\ }\textbf {\bibinfo {volume} {455}},\ \bibinfo {pages} {648}
  (\bibinfo {year} {2008})}\BibitemShut {NoStop}%
\bibitem [{\citenamefont {Pervushin}\ \emph {et~al.}(1997)\citenamefont
  {Pervushin}, \citenamefont {Riek}, \citenamefont {Wider},\ and\ \citenamefont
  {W{\"u}thrich}}]{pervushin1997attenuated}%
  \BibitemOpen
  \bibfield  {author} {\bibinfo {author} {\bibfnamefont {K.}~\bibnamefont
  {Pervushin}}, \bibinfo {author} {\bibfnamefont {R.}~\bibnamefont {Riek}},
  \bibinfo {author} {\bibfnamefont {G.}~\bibnamefont {Wider}}, \ and\ \bibinfo
  {author} {\bibfnamefont {K.}~\bibnamefont {W{\"u}thrich}},\ }\href@noop {}
  {\bibfield  {journal} {\bibinfo  {journal} {Proceedings of the National
  Academy of Sciences}\ }\textbf {\bibinfo {volume} {94}},\ \bibinfo {pages}
  {12366} (\bibinfo {year} {1997})}\BibitemShut {NoStop}%
\bibitem [{\citenamefont {Yang}\ \emph {et~al.}(2010)\citenamefont {Yang},
  \citenamefont {Xu}, \citenamefont {Song},\ and\ \citenamefont
  {Sun}}]{yang2010dimerization}%
  \BibitemOpen
  \bibfield  {author} {\bibinfo {author} {\bibfnamefont {S.}~\bibnamefont
  {Yang}}, \bibinfo {author} {\bibfnamefont {D.}~\bibnamefont {Xu}}, \bibinfo
  {author} {\bibfnamefont {Z.}~\bibnamefont {Song}}, \ and\ \bibinfo {author}
  {\bibfnamefont {C.}~\bibnamefont {Sun}},\ }\href@noop {} {\bibfield
  {journal} {\bibinfo  {journal} {The Journal of chemical physics}\ }\textbf
  {\bibinfo {volume} {132}},\ \bibinfo {pages} {234501} (\bibinfo {year}
  {2010})}\BibitemShut {NoStop}%
\bibitem [{\citenamefont {El-Ganainy}\ and\ \citenamefont
  {John}(2013)}]{el2013resonant}%
  \BibitemOpen
  \bibfield  {author} {\bibinfo {author} {\bibfnamefont {R.}~\bibnamefont
  {El-Ganainy}}\ and\ \bibinfo {author} {\bibfnamefont {S.}~\bibnamefont
  {John}},\ }\href@noop {} {\bibfield  {journal} {\bibinfo  {journal} {New
  Journal of Physics}\ }\textbf {\bibinfo {volume} {15}},\ \bibinfo {pages}
  {083033} (\bibinfo {year} {2013})}\BibitemShut {NoStop}%
\bibitem [{\citenamefont {Martinis}\ \emph
  {et~al.}(2005{\natexlab{a}})\citenamefont {Martinis}, \citenamefont {Cooper},
  \citenamefont {McDermott}, \citenamefont {Steffen}, \citenamefont {Ansmann},
  \citenamefont {Osborn}, \citenamefont {Cicak}, \citenamefont {Oh},
  \citenamefont {Pappas}, \citenamefont {Simmonds} \emph
  {et~al.}}]{martinis2005decoherence}%
  \BibitemOpen
  \bibfield  {author} {\bibinfo {author} {\bibfnamefont {J.~M.}\ \bibnamefont
  {Martinis}}, \bibinfo {author} {\bibfnamefont {K.~B.}\ \bibnamefont
  {Cooper}}, \bibinfo {author} {\bibfnamefont {R.}~\bibnamefont {McDermott}},
  \bibinfo {author} {\bibfnamefont {M.}~\bibnamefont {Steffen}}, \bibinfo
  {author} {\bibfnamefont {M.}~\bibnamefont {Ansmann}}, \bibinfo {author}
  {\bibfnamefont {K.}~\bibnamefont {Osborn}}, \bibinfo {author} {\bibfnamefont
  {K.}~\bibnamefont {Cicak}}, \bibinfo {author} {\bibfnamefont
  {S.}~\bibnamefont {Oh}}, \bibinfo {author} {\bibfnamefont {D.~P.}\
  \bibnamefont {Pappas}}, \bibinfo {author} {\bibfnamefont {R.~W.}\
  \bibnamefont {Simmonds}},  \emph {et~al.},\ }\href@noop {} {\bibfield
  {journal} {\bibinfo  {journal} {Physical Review Letters}\ }\textbf {\bibinfo
  {volume} {95}},\ \bibinfo {pages} {210503} (\bibinfo {year}
  {2005}{\natexlab{a}})}\BibitemShut {NoStop}%
\bibitem [{\citenamefont {Paik}\ \emph {et~al.}(2011)\citenamefont {Paik},
  \citenamefont {Schuster}, \citenamefont {Bishop}, \citenamefont {Kirchmair},
  \citenamefont {Catelani}, \citenamefont {Sears}, \citenamefont {Johnson},
  \citenamefont {Reagor}, \citenamefont {Frunzio}, \citenamefont {Glazman}
  \emph {et~al.}}]{paik2011observation}%
  \BibitemOpen
  \bibfield  {author} {\bibinfo {author} {\bibfnamefont {H.}~\bibnamefont
  {Paik}}, \bibinfo {author} {\bibfnamefont {D.}~\bibnamefont {Schuster}},
  \bibinfo {author} {\bibfnamefont {L.~S.}\ \bibnamefont {Bishop}}, \bibinfo
  {author} {\bibfnamefont {G.}~\bibnamefont {Kirchmair}}, \bibinfo {author}
  {\bibfnamefont {G.}~\bibnamefont {Catelani}}, \bibinfo {author}
  {\bibfnamefont {A.}~\bibnamefont {Sears}}, \bibinfo {author} {\bibfnamefont
  {B.}~\bibnamefont {Johnson}}, \bibinfo {author} {\bibfnamefont
  {M.}~\bibnamefont {Reagor}}, \bibinfo {author} {\bibfnamefont
  {L.}~\bibnamefont {Frunzio}}, \bibinfo {author} {\bibfnamefont
  {L.}~\bibnamefont {Glazman}},  \emph {et~al.},\ }\href@noop {} {\bibfield
  {journal} {\bibinfo  {journal} {Physical Review Letters}\ }\textbf {\bibinfo
  {volume} {107}},\ \bibinfo {pages} {240501} (\bibinfo {year}
  {2011})}\BibitemShut {NoStop}%
\bibitem [{\citenamefont {Rigetti}\ \emph {et~al.}(2012)\citenamefont
  {Rigetti}, \citenamefont {Gambetta}, \citenamefont {Poletto}, \citenamefont
  {Plourde}, \citenamefont {Chow}, \citenamefont {C{\'o}rcoles}, \citenamefont
  {Smolin}, \citenamefont {Merkel}, \citenamefont {Rozen}, \citenamefont
  {Keefe} \emph {et~al.}}]{rigetti2012superconducting}%
  \BibitemOpen
  \bibfield  {author} {\bibinfo {author} {\bibfnamefont {C.}~\bibnamefont
  {Rigetti}}, \bibinfo {author} {\bibfnamefont {J.~M.}\ \bibnamefont
  {Gambetta}}, \bibinfo {author} {\bibfnamefont {S.}~\bibnamefont {Poletto}},
  \bibinfo {author} {\bibfnamefont {B.}~\bibnamefont {Plourde}}, \bibinfo
  {author} {\bibfnamefont {J.~M.}\ \bibnamefont {Chow}}, \bibinfo {author}
  {\bibfnamefont {A.}~\bibnamefont {C{\'o}rcoles}}, \bibinfo {author}
  {\bibfnamefont {J.~A.}\ \bibnamefont {Smolin}}, \bibinfo {author}
  {\bibfnamefont {S.~T.}\ \bibnamefont {Merkel}}, \bibinfo {author}
  {\bibfnamefont {J.}~\bibnamefont {Rozen}}, \bibinfo {author} {\bibfnamefont
  {G.~A.}\ \bibnamefont {Keefe}},  \emph {et~al.},\ }\href@noop {} {\bibfield
  {journal} {\bibinfo  {journal} {Physical Review B}\ }\textbf {\bibinfo
  {volume} {86}},\ \bibinfo {pages} {100506} (\bibinfo {year}
  {2012})}\BibitemShut {NoStop}%
\bibitem [{\citenamefont {Lisenfeld}\ \emph {et~al.}(2016)\citenamefont
  {Lisenfeld}, \citenamefont {Bilmes}, \citenamefont {Matityahu}, \citenamefont
  {Zanker}, \citenamefont {Marthaler}, \citenamefont {Schechter}, \citenamefont
  {Sch{\"o}n}, \citenamefont {Shnirman}, \citenamefont {Weiss},\ and\
  \citenamefont {Ustinov}}]{lisenfeld_decoherence_2016}%
  \BibitemOpen
  \bibfield  {author} {\bibinfo {author} {\bibfnamefont {J.}~\bibnamefont
  {Lisenfeld}}, \bibinfo {author} {\bibfnamefont {A.}~\bibnamefont {Bilmes}},
  \bibinfo {author} {\bibfnamefont {S.}~\bibnamefont {Matityahu}}, \bibinfo
  {author} {\bibfnamefont {S.}~\bibnamefont {Zanker}}, \bibinfo {author}
  {\bibfnamefont {M.}~\bibnamefont {Marthaler}}, \bibinfo {author}
  {\bibfnamefont {M.}~\bibnamefont {Schechter}}, \bibinfo {author}
  {\bibfnamefont {G.}~\bibnamefont {Sch{\"o}n}}, \bibinfo {author}
  {\bibfnamefont {A.}~\bibnamefont {Shnirman}}, \bibinfo {author}
  {\bibfnamefont {G.}~\bibnamefont {Weiss}}, \ and\ \bibinfo {author}
  {\bibfnamefont {A.~V.}\ \bibnamefont {Ustinov}},\ }\href {\doibase
  10.1038/srep23786} {\bibfield  {journal} {\bibinfo  {journal} {Sci. Rep.}\
  }\textbf {\bibinfo {volume} {6}},\ \bibinfo {pages} {23786} (\bibinfo {year}
  {2016})}\BibitemShut {NoStop}%
\bibitem [{\citenamefont {Martinis}\ \emph
  {et~al.}(2005{\natexlab{b}})\citenamefont {Martinis}, \citenamefont {Cooper},
  \citenamefont {McDermott}, \citenamefont {Steffen}, \citenamefont {Ansmann},
  \citenamefont {Osborn}, \citenamefont {Cicak}, \citenamefont {Oh},
  \citenamefont {Pappas}, \citenamefont {Simmonds},\ and\ \citenamefont
  {Yu}}]{martinis_decoherence_2005}%
  \BibitemOpen
  \bibfield  {author} {\bibinfo {author} {\bibfnamefont {J.~M.}\ \bibnamefont
  {Martinis}}, \bibinfo {author} {\bibfnamefont {K.~B.}\ \bibnamefont
  {Cooper}}, \bibinfo {author} {\bibfnamefont {R.}~\bibnamefont {McDermott}},
  \bibinfo {author} {\bibfnamefont {M.}~\bibnamefont {Steffen}}, \bibinfo
  {author} {\bibfnamefont {M.}~\bibnamefont {Ansmann}}, \bibinfo {author}
  {\bibfnamefont {K.~D.}\ \bibnamefont {Osborn}}, \bibinfo {author}
  {\bibfnamefont {K.}~\bibnamefont {Cicak}}, \bibinfo {author} {\bibfnamefont
  {S.}~\bibnamefont {Oh}}, \bibinfo {author} {\bibfnamefont {D.~P.}\
  \bibnamefont {Pappas}}, \bibinfo {author} {\bibfnamefont {R.~W.}\
  \bibnamefont {Simmonds}}, \ and\ \bibinfo {author} {\bibfnamefont {C.~C.}\
  \bibnamefont {Yu}},\ }\href {\doibase 10.1103/PhysRevLett.95.210503}
  {\bibfield  {journal} {\bibinfo  {journal} {Phys. Rev. Lett.}\ }\textbf
  {\bibinfo {volume} {95}},\ \bibinfo {pages} {210503} (\bibinfo {year}
  {2005}{\natexlab{b}})}\BibitemShut {NoStop}%
\bibitem [{\citenamefont {Zhao}\ \emph {et~al.}(2012)\citenamefont {Zhao},
  \citenamefont {Ho},\ and\ \citenamefont {Liu}}]{zhao2012decoherence}%
  \BibitemOpen
  \bibfield  {author} {\bibinfo {author} {\bibfnamefont {N.}~\bibnamefont
  {Zhao}}, \bibinfo {author} {\bibfnamefont {S.-W.}\ \bibnamefont {Ho}}, \ and\
  \bibinfo {author} {\bibfnamefont {R.-B.}\ \bibnamefont {Liu}},\ }\href@noop
  {} {\bibfield  {journal} {\bibinfo  {journal} {Physical Review B}\ }\textbf
  {\bibinfo {volume} {85}},\ \bibinfo {pages} {115303} (\bibinfo {year}
  {2012})}\BibitemShut {NoStop}%
\bibitem [{\citenamefont {{\v{Z}}nidari{\v{c}}}\ \emph
  {et~al.}(2008)\citenamefont {{\v{Z}}nidari{\v{c}}}, \citenamefont {Prosen},\
  and\ \citenamefont {Prelov{\v{s}}ek}}]{vznidarivc2008many}%
  \BibitemOpen
  \bibfield  {author} {\bibinfo {author} {\bibfnamefont {M.}~\bibnamefont
  {{\v{Z}}nidari{\v{c}}}}, \bibinfo {author} {\bibfnamefont {T.}~\bibnamefont
  {Prosen}}, \ and\ \bibinfo {author} {\bibfnamefont {P.}~\bibnamefont
  {Prelov{\v{s}}ek}},\ }\href@noop {} {\bibfield  {journal} {\bibinfo
  {journal} {Physical Review B}\ }\textbf {\bibinfo {volume} {77}},\ \bibinfo
  {pages} {064426} (\bibinfo {year} {2008})}\BibitemShut {NoStop}%
\bibitem [{\citenamefont {Doherty}\ \emph {et~al.}(2013)\citenamefont
  {Doherty}, \citenamefont {Manson}, \citenamefont {Delaney}, \citenamefont
  {Jelezko}, \citenamefont {Wrachtrup},\ and\ \citenamefont
  {Hollenberg}}]{doherty_nitrogen-vacancy_2013}%
  \BibitemOpen
  \bibfield  {author} {\bibinfo {author} {\bibfnamefont {M.~W.}\ \bibnamefont
  {Doherty}}, \bibinfo {author} {\bibfnamefont {N.~B.}\ \bibnamefont {Manson}},
  \bibinfo {author} {\bibfnamefont {P.}~\bibnamefont {Delaney}}, \bibinfo
  {author} {\bibfnamefont {F.}~\bibnamefont {Jelezko}}, \bibinfo {author}
  {\bibfnamefont {J.}~\bibnamefont {Wrachtrup}}, \ and\ \bibinfo {author}
  {\bibfnamefont {L.~C.~L.}\ \bibnamefont {Hollenberg}},\ }\href {\doibase
  10.1016/j.physrep.2013.02.001} {\bibfield  {journal} {\bibinfo  {journal}
  {Phys. Rep.}\ }\textbf {\bibinfo {volume} {528}},\ \bibinfo {pages} {1}
  (\bibinfo {year} {2013})}\BibitemShut {NoStop}%
\bibitem [{\citenamefont {Wrachtrup}\ and\ \citenamefont
  {Jelezko}(2006)}]{wrachtrup_processing_2006}%
  \BibitemOpen
  \bibfield  {author} {\bibinfo {author} {\bibfnamefont {J.}~\bibnamefont
  {Wrachtrup}}\ and\ \bibinfo {author} {\bibfnamefont {F.}~\bibnamefont
  {Jelezko}},\ }\href {\doibase 10.1088/0953-8984/18/21/S08} {\bibfield
  {journal} {\bibinfo  {journal} {J. Phys.: Condens. Matter}\ }\textbf
  {\bibinfo {volume} {18}},\ \bibinfo {pages} {S807} (\bibinfo {year}
  {2006})}\BibitemShut {NoStop}%
\bibitem [{\citenamefont {Jackson}(1998)}]{jackson_classical_1998}%
  \BibitemOpen
  \bibfield  {author} {\bibinfo {author} {\bibfnamefont {J.~D.}\ \bibnamefont
  {Jackson}},\ }\href@noop {} {\emph {\bibinfo {title} {Classical
  {Electrodynamics} {Third} {Edition}}}},\ \bibinfo {edition} {3rd}\ ed.\
  (\bibinfo  {publisher} {Wiley},\ \bibinfo {address} {New York},\ \bibinfo
  {year} {1998})\BibitemShut {NoStop}%
\bibitem [{\citenamefont {Lehmberg}(1970)}]{lehmberg1970radiation}%
  \BibitemOpen
  \bibfield  {author} {\bibinfo {author} {\bibfnamefont {R.}~\bibnamefont
  {Lehmberg}},\ }\href@noop {} {\bibfield  {journal} {\bibinfo  {journal}
  {Physical Review A}\ }\textbf {\bibinfo {volume} {2}},\ \bibinfo {pages}
  {883} (\bibinfo {year} {1970})}\BibitemShut {NoStop}%
\bibitem [{\citenamefont {Agarwal}(1974)}]{agarwal_quantum_1974}%
  \BibitemOpen
  \bibfield  {author} {\bibinfo {author} {\bibfnamefont {G.~S.}\ \bibnamefont
  {Agarwal}},\ }\href@noop {} {\emph {\bibinfo {title} {Quantum statistical
  theories of spontaneous emission and their relation to other approaches}}}\
  (\bibinfo  {publisher} {Springer},\ \bibinfo {year} {1974})\BibitemShut
  {NoStop}%
\bibitem [{\citenamefont {Ficek}\ \emph {et~al.}(1987)\citenamefont {Ficek},
  \citenamefont {Tana{\'s}},\ and\ \citenamefont {Kielich}}]{ficek1987quantum}%
  \BibitemOpen
  \bibfield  {author} {\bibinfo {author} {\bibfnamefont {Z.}~\bibnamefont
  {Ficek}}, \bibinfo {author} {\bibfnamefont {R.}~\bibnamefont {Tana{\'s}}}, \
  and\ \bibinfo {author} {\bibfnamefont {S.}~\bibnamefont {Kielich}},\
  }\href@noop {} {\bibfield  {journal} {\bibinfo  {journal} {Physica A:
  Statistical Mechanics and its Applications}\ }\textbf {\bibinfo {volume}
  {146}},\ \bibinfo {pages} {452} (\bibinfo {year} {1987})}\BibitemShut
  {NoStop}%
\bibitem [{\citenamefont {Craig}\ and\ \citenamefont
  {Thirunamachandran}(1998)}]{craig_molecular_1998}%
  \BibitemOpen
  \bibfield  {author} {\bibinfo {author} {\bibfnamefont {D.~P.}\ \bibnamefont
  {Craig}}\ and\ \bibinfo {author} {\bibfnamefont {T.}~\bibnamefont
  {Thirunamachandran}},\ }\href@noop {} {\emph {\bibinfo {title} {Molecular
  quantum electrodynamics}}}\ (\bibinfo  {publisher} {Dover Publications},\
  \bibinfo {address} {Mineola, N.Y},\ \bibinfo {year} {1998})\BibitemShut
  {NoStop}%
\bibitem [{\citenamefont {Ficek}\ and\ \citenamefont
  {Swain}(2005)}]{ficek2005quantum}%
  \BibitemOpen
  \bibfield  {author} {\bibinfo {author} {\bibfnamefont {Z.}~\bibnamefont
  {Ficek}}\ and\ \bibinfo {author} {\bibfnamefont {S.}~\bibnamefont {Swain}},\
  }\href@noop {} {\emph {\bibinfo {title} {Quantum interference and coherence:
  theory and experiments}}},\ Vol.\ \bibinfo {volume} {100}\ (\bibinfo
  {publisher} {Springer Science \& Business Media},\ \bibinfo {year}
  {2005})\BibitemShut {NoStop}%
\bibitem [{\citenamefont {Dicke}(1954)}]{dicke1954coherence}%
  \BibitemOpen
  \bibfield  {author} {\bibinfo {author} {\bibfnamefont {R.~H.}\ \bibnamefont
  {Dicke}},\ }\href@noop {} {\bibfield  {journal} {\bibinfo  {journal}
  {Physical Review}\ }\textbf {\bibinfo {volume} {93}},\ \bibinfo {pages} {99}
  (\bibinfo {year} {1954})}\BibitemShut {NoStop}%
\bibitem [{\citenamefont {Wang}\ \emph {et~al.}(2018)\citenamefont {Wang},
  \citenamefont {Dong},\ and\ \citenamefont {Li}}]{wang2018magnetic}%
  \BibitemOpen
  \bibfield  {author} {\bibinfo {author} {\bibfnamefont {J.}~\bibnamefont
  {Wang}}, \bibinfo {author} {\bibfnamefont {H.}~\bibnamefont {Dong}}, \ and\
  \bibinfo {author} {\bibfnamefont {S.-W.}\ \bibnamefont {Li}},\ }\href@noop {}
  {\bibfield  {journal} {\bibinfo  {journal} {Physical Review A}\ }\textbf
  {\bibinfo {volume} {97}},\ \bibinfo {pages} {013819} (\bibinfo {year}
  {2018})}\BibitemShut {NoStop}%
\bibitem [{\citenamefont {Cohen-Tannoudji}\ \emph {et~al.}(1989)\citenamefont
  {Cohen-Tannoudji}, \citenamefont {Dupont-Roc},\ and\ \citenamefont
  {Grynberg}}]{cohen-tannoudji_photons_1989}%
  \BibitemOpen
  \bibfield  {author} {\bibinfo {author} {\bibfnamefont {C.}~\bibnamefont
  {Cohen-Tannoudji}}, \bibinfo {author} {\bibfnamefont {J.}~\bibnamefont
  {Dupont-Roc}}, \ and\ \bibinfo {author} {\bibfnamefont {G.}~\bibnamefont
  {Grynberg}},\ }\href@noop {} {\emph {\bibinfo {title} {Photons and {Atoms}:
  {Introduction} to {Quantum} {Electrodynamics}}}},\ \bibinfo {edition} {1st}\
  ed.\ (\bibinfo  {publisher} {Wiley-VCH},\ \bibinfo {address} {New York},\
  \bibinfo {year} {1989})\BibitemShut {NoStop}%
\bibitem [{\citenamefont {Weinberg}(2012)}]{weinberg2012lectures}%
  \BibitemOpen
  \bibfield  {author} {\bibinfo {author} {\bibfnamefont {S.}~\bibnamefont
  {Weinberg}},\ }\href@noop {} {\bibfield  {journal} {\bibinfo  {journal}
  {Lectures on Quantum Mechanics}\ } (\bibinfo {year} {Cambridge University
  Press, Cambridge, UK, 2012})}\BibitemShut {NoStop}%
\bibitem [{\citenamefont {Breuer}\ and\ \citenamefont
  {Petruccione}(2002)}]{breuer_theory_2002}%
  \BibitemOpen
  \bibfield  {author} {\bibinfo {author} {\bibfnamefont {H.}~\bibnamefont
  {Breuer}}\ and\ \bibinfo {author} {\bibfnamefont {F.}~\bibnamefont
  {Petruccione}},\ }\href@noop {} {\emph {\bibinfo {title} {The theory of open
  quantum systems}}}\ (\bibinfo  {publisher} {Oxford University Press},\
  \bibinfo {year} {2002})\BibitemShut {NoStop}%
\bibitem [{\citenamefont {Li}\ \emph {et~al.}(2015)\citenamefont {Li},
  \citenamefont {Cai},\ and\ \citenamefont {Sun}}]{li2015steady}%
  \BibitemOpen
  \bibfield  {author} {\bibinfo {author} {\bibfnamefont {S.-W.}\ \bibnamefont
  {Li}}, \bibinfo {author} {\bibfnamefont {C.}~\bibnamefont {Cai}}, \ and\
  \bibinfo {author} {\bibfnamefont {C.}~\bibnamefont {Sun}},\ }\href@noop {}
  {\bibfield  {journal} {\bibinfo  {journal} {Annals of Physics}\ }\textbf
  {\bibinfo {volume} {360}},\ \bibinfo {pages} {19} (\bibinfo {year}
  {2015})}\BibitemShut {NoStop}%
\bibitem [{\citenamefont {Li}\ \emph {et~al.}(2016)\citenamefont {Li},
  \citenamefont {Kim},\ and\ \citenamefont {Scully}}]{li2016non}%
  \BibitemOpen
  \bibfield  {author} {\bibinfo {author} {\bibfnamefont {S.-W.}\ \bibnamefont
  {Li}}, \bibinfo {author} {\bibfnamefont {M.~B.}\ \bibnamefont {Kim}}, \ and\
  \bibinfo {author} {\bibfnamefont {M.~O.}\ \bibnamefont {Scully}},\
  }\href@noop {} {\bibfield  {journal} {\bibinfo  {journal} {arXiv preprint
  arXiv:1604.03091}\ } (\bibinfo {year} {2016})}\BibitemShut {NoStop}%
\bibitem [{\citenamefont {Baranov}\ \emph {et~al.}(2016)\citenamefont
  {Baranov}, \citenamefont {Savelev}, \citenamefont {Li}, \citenamefont
  {Krasnok},\ and\ \citenamefont {Al{\`u}}}]{baranov_magnetic_2016}%
  \BibitemOpen
  \bibfield  {author} {\bibinfo {author} {\bibfnamefont {D.~G.}\ \bibnamefont
  {Baranov}}, \bibinfo {author} {\bibfnamefont {R.~S.}\ \bibnamefont
  {Savelev}}, \bibinfo {author} {\bibfnamefont {S.~V.}\ \bibnamefont {Li}},
  \bibinfo {author} {\bibfnamefont {A.~E.}\ \bibnamefont {Krasnok}}, \ and\
  \bibinfo {author} {\bibfnamefont {A.}~\bibnamefont {Al{\`u}}},\ }\href
  {http://arxiv.org/abs/1610.02001} {\bibfield  {journal} {\bibinfo  {journal}
  {arXiv:1610.02001}\ } (\bibinfo {year} {2016})}\BibitemShut {NoStop}%
\bibitem [{\citenamefont {Donaire}\ \emph {et~al.}(2017)\citenamefont
  {Donaire}, \citenamefont {Mu{\~n}oz-Casta{\~n}eda},\ and\ \citenamefont
  {Nieto}}]{donaire_dipole-dipole_2017}%
  \BibitemOpen
  \bibfield  {author} {\bibinfo {author} {\bibfnamefont {M.}~\bibnamefont
  {Donaire}}, \bibinfo {author} {\bibfnamefont {J.~M.}\ \bibnamefont
  {Mu{\~n}oz-Casta{\~n}eda}}, \ and\ \bibinfo {author} {\bibfnamefont {L.~M.}\
  \bibnamefont {Nieto}},\ }\href {\doibase 10.1103/PhysRevA.96.042714}
  {\bibfield  {journal} {\bibinfo  {journal} {Phys. Rev. A}\ }\textbf {\bibinfo
  {volume} {96}},\ \bibinfo {pages} {042714} (\bibinfo {year}
  {2017})}\BibitemShut {NoStop}%
\bibitem [{\citenamefont {Tang}\ \emph {et~al.}(2018)\citenamefont {Tang},
  \citenamefont {Kao}, \citenamefont {Li},\ and\ \citenamefont
  {Lev}}]{tang_tuning_2018}%
  \BibitemOpen
  \bibfield  {author} {\bibinfo {author} {\bibfnamefont {Y.}~\bibnamefont
  {Tang}}, \bibinfo {author} {\bibfnamefont {W.}~\bibnamefont {Kao}}, \bibinfo
  {author} {\bibfnamefont {K.-Y.}\ \bibnamefont {Li}}, \ and\ \bibinfo {author}
  {\bibfnamefont {B.~L.}\ \bibnamefont {Lev}},\ }\href {\doibase
  10.1103/PhysRevLett.120.230401} {\bibfield  {journal} {\bibinfo  {journal}
  {Phys. Rev. Lett.}\ }\textbf {\bibinfo {volume} {120}} (\bibinfo {year}
  {2018}),\ 10.1103/PhysRevLett.120.230401}\BibitemShut {NoStop}%
\bibitem [{\citenamefont {Davis}\ \emph {et~al.}(2019)\citenamefont {Davis},
  \citenamefont {Bentsen}, \citenamefont {Homeier}, \citenamefont {Li},\ and\
  \citenamefont {Schleier-Smith}}]{davis_photon-mediated_2019}%
  \BibitemOpen
  \bibfield  {author} {\bibinfo {author} {\bibfnamefont {E.~J.}\ \bibnamefont
  {Davis}}, \bibinfo {author} {\bibfnamefont {G.}~\bibnamefont {Bentsen}},
  \bibinfo {author} {\bibfnamefont {L.}~\bibnamefont {Homeier}}, \bibinfo
  {author} {\bibfnamefont {T.}~\bibnamefont {Li}}, \ and\ \bibinfo {author}
  {\bibfnamefont {M.~H.}\ \bibnamefont {Schleier-Smith}},\ }\href {\doibase
  10.1103/PhysRevLett.122.010405} {\bibfield  {journal} {\bibinfo  {journal}
  {Phys. Rev. Lett.}\ }\textbf {\bibinfo {volume} {122}} (\bibinfo {year}
  {2019}),\ 10.1103/PhysRevLett.122.010405}\BibitemShut {NoStop}%
\end{thebibliography}%
 
\end{document}